\documentclass[fleqn,aps,prc,onecolumn,nofootinbib,letterpaper,superscriptaddress]{revtex4}
\usepackage{dcolumn}
\usepackage{color}
\usepackage{graphicx}
\usepackage[colorlinks=true]{hyperref}
\usepackage{amsmath}
\usepackage{multirow}
\usepackage{bm}
\usepackage{float}
\usepackage{lineno}
\usepackage{caption}
\usepackage{subfigure}

\begin{document}

\title{ $^{83}$Rb/$^{83\rm{m}}$Kr production and cross-section measurement with 3.4~MeV and 20~MeV proton beams}
\begin{abstract}
  $\rm ^{83m}Kr$, with a short lifetime, is an ideal calibration source for liquid xenon or liquid argon detectors. The $\rm ^{83m}Kr$ isomer can be generated through the decay of $\rm ^{83} Rb$ isotope which is usually produced by proton beams bombarding natural krypton atoms. In this paper, we report a successful production of $\rm ^{83}Rb/^{83m}Kr$
  with a proton beam energy of 3.4 MeV, and the first measurement of the production rate with such low energy proton beams. Another production attempt is performed using the newly available 20 MeV proton beam in China, and the measured production rate is consistent with previous measurements. The produced $\rm ^{83m}Kr$ source has been successfully injected into the PandaX-II liquid xenon detector, yielding enough statistics for detector calibration.
\end{abstract}

\def\sjtu{School of Physics and Astronomy, Shanghai Jiao Tong University, MOE Key Laboratory for Particle Astrophysics and Cosmology, Shanghai Key Laboratory for Particle Physics and Cosmology, Shanghai 200240, China}
\def\md{Department of Physics, University of Maryland, College Park, Maryland 20742, USA}
\def\zhiyuan{Zhiyuan College, Shanghai Jiao Tong University, Shanghai 200240, China}
\def\tdli{Tsung-Dao Lee Institute, Shanghai Jiao Tong University, Shanghai 200240, China}
\def\sjtusc{Shanghai Jiao Tong University Sichuan Research Institute, Chengdu 610213, China}
\def\ciae{China Institute of Atomic Energy, Beijing 102413, China}
\def\imp{Institute of Modern Physics, the Chinese Academy of Sciences, Lanzhou 730000, China}
\def\ihep{Institute of High Energy Physics, the Chinese Academy of Sciences, Beijing 100049, China}
\def\fd{Institute of Modern Physics, Fudan University, Shanghai 200433, China}

\affiliation{\sjtu}
\author{Dan Zhang}\affiliation{\md}
\author{Yifan Li}\affiliation{\zhiyuan}
\author{Jie Bao}\affiliation{\ciae}
\author{Changbo Fu}\affiliation{\fd}
\author{Mengyun Guan}\affiliation{\ihep}
\author{Yuan He}\affiliation{\imp}
\author{Xiangdong Ji}\affiliation{\md}
\author{Huan Jia}\affiliation{\imp}
\author{Yao Li}\affiliation{\sjtu}
\author{Jianglai Liu}\affiliation{\sjtu}\affiliation{\tdli}\affiliation{\sjtusc}
\author{Jingkai Xia}\email[Corresponding author: ]{xiajk@sjtu.edu.cn}\email[Now at ShanghaiTech University, Shanghai, China]{}\affiliation{\sjtu}
\author{Weixing Xiong}\affiliation{\ihep}
\author{Jingtao You}\affiliation{\zhiyuan}
\author{Chenzhang Yuan}\affiliation{\imp}
\author{Ning Zhou}\email[Corresponding author: ]{nzhou@sjtu.edu.cn}\affiliation{\sjtu}

\maketitle

\section{Introduction}
Dark matter searches in xenon detectors such as PandaX-II~\cite{pandax2nr,pandax2er} and its successors~\cite{pandax4t,karl} rely on accurate event reconstruction from scintillation (S1) and ionization (S2) signals generated by energy deposit in liquid xenon. Due to detector geometric deformation and electric field non-uniformity, the magnitudes of S1 and S2 signals have strong position dependence, which degrades the event reconstruction resolution and the discrimination between nuclear recoil and electron recoil events~\cite{3dmodel}. Therefore, it is necessary to have mono-energetic signals evenly distributed in the detector in order to calibrate the detector response. 

One such calibration source is the $^{83m}$Kr isomer, which is gaseous and can be mixed with xenon uniformly. Its half-life is only 1.83~h, so no specific removal procedure is required after a calibration campaign. In addition, the energy of the $^{83m}$Kr decays is small enough to calibrate noble-liquid detectors for dark matter searches where the region of interest is usually less than 100~keV~\cite{luxkr,vhannen,arkr}.
$^{83m}$Kr has been used to calibrate tritium $\beta$ decay experiments~\cite{katrin}, calorimeters in the large electron-positron colliders~\cite{ce83mkr} and the heavy ion detector of the ALICE experiment~\cite{alice}. An experiment at Yale University reported the first use of $^{83m}$Kr in liquid noble element detectors for spatial and energy calibration~\cite{yale, yaleNeon}. 

Because $^{83m}$Kr is a short-living source, we resort to its mother isotope $^{83}$Rb for source preparation, which has a relatively long lifetime ($T_{1/2}=86.2$~days).
The full decay scheme of $^{83}$Rb according to NNDC
shows that 74.385\% of the $^{83}$Rb atoms decay into the $^{83m}$Kr isomer~\cite{nndc}. The simplicity of the decay mode mitigates potential side effects for detector calibration. 

$^{83}$Rb is a synthetic radioisotope that can be produced by proton beams bombarding natural krypton with peak production rate at around 20~MeV proton energy. Due to limited access to such a high energy proton facility, we tested the production with a lower energy proton beam. In this paper, we report a successful production of $^{83}$Rb/$^{83m}$Kr with 3.4~MeV proton beam at the China Institute of Atomic Energy, and the first measurement of the yield of the $\rm ^{nat}$Kr$\rm(p,xn)^{83}$Rb reaction for proton energy below 5~MeV. The bombarding chamber design and experiment setup are demonstrated in section~\ref{sec:setup}. The collection of $^{83}$Rb/$^{83m}$Kr product and the measurement of production rate are presented in section~\ref{sec:measurement}. Another production test performed with a recently available 20~MeV proton beam at the Institute of Modern Physics, Chinese Academy of Sciences is shown in section~\ref{sec:production20mev}. Finally, the injection test of $^{83}$Rb/$^{83m}$Kr in the PandaX-II detector is described in section~\ref{sec:injection}. 

\section{Production of $^{83}$Rb/${}^{83m}$Kr source with 3.4~MeV protons}
\label{sec:setup}
$^{83}$Rb can be synthesized by the bombardment of krypton with protons or bromine with $\alpha$ particles.  Since $\alpha$ particle bombarding bromine yields more unexpected isotopes which might contaminate our detector, the $^{\rm{nat}}$Kr(p, xn)$^{83}$Rb process is favored for our calibration purpose. The production rate depends strongly on the energy of the proton beam. The optimal energy is approximately 20 MeV, which can maximize the $^{83}$Rb production rate and minimize the unwanted $^{84}$Rb and $^{86}$Rb that bring an extra risk of increasing backgrounds in the gamma spectra~\cite{optim}. 

Before we had access to the 20~MeV proton beams, the China Institute of Atomic Energy provided proton beams with energy approximately 3.4~MeV ({\color{black}NEC 1.7 MV Model 5SDH-2 Tandem Accelerator}), which is slightly above the theoretical energy threshold for the $^{83}$Kr(p,n)$^{83}$Rb reaction ($E_{\rm th} = 1.7$~MeV). $^{86}$Rb isotope ($T_{1/2}=18.6$~d) is also expected during production as the $^{86}$Kr(p,n)$^{86}$Rb reaction has an energy threshold of 1.3~MeV.

In this work, two $^{\rm{nat}}$Kr target cells are prepared for bombardment so that one can be used to study the outcome in detail and the other can be supplied as a calibration source. The cell design is shown in Fig.~\ref{fig:chamberdesign}, details are described as following:
\begin{figure}[htbp]
\centering
\includegraphics[width=4.5 in]{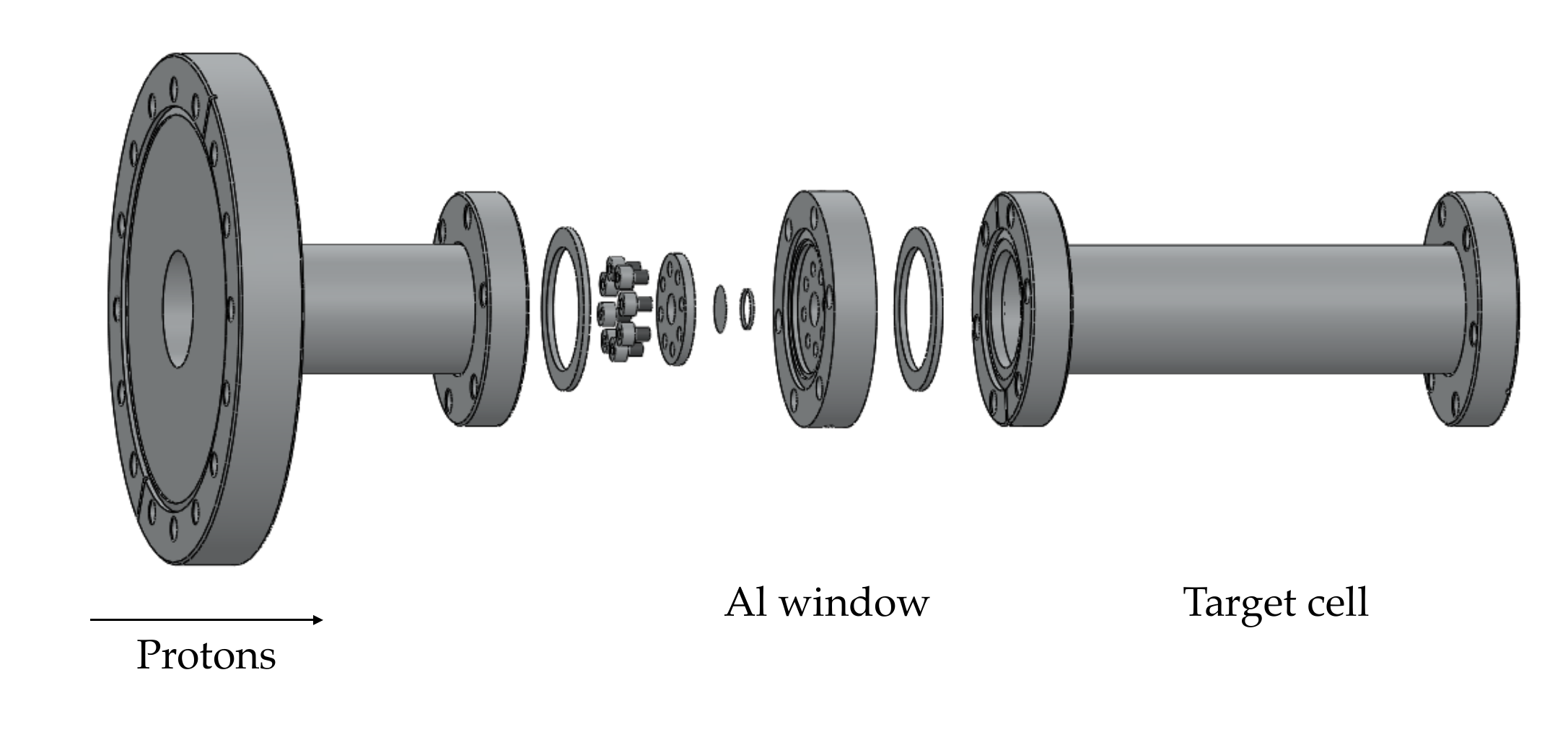}
\includegraphics[width=3. in]{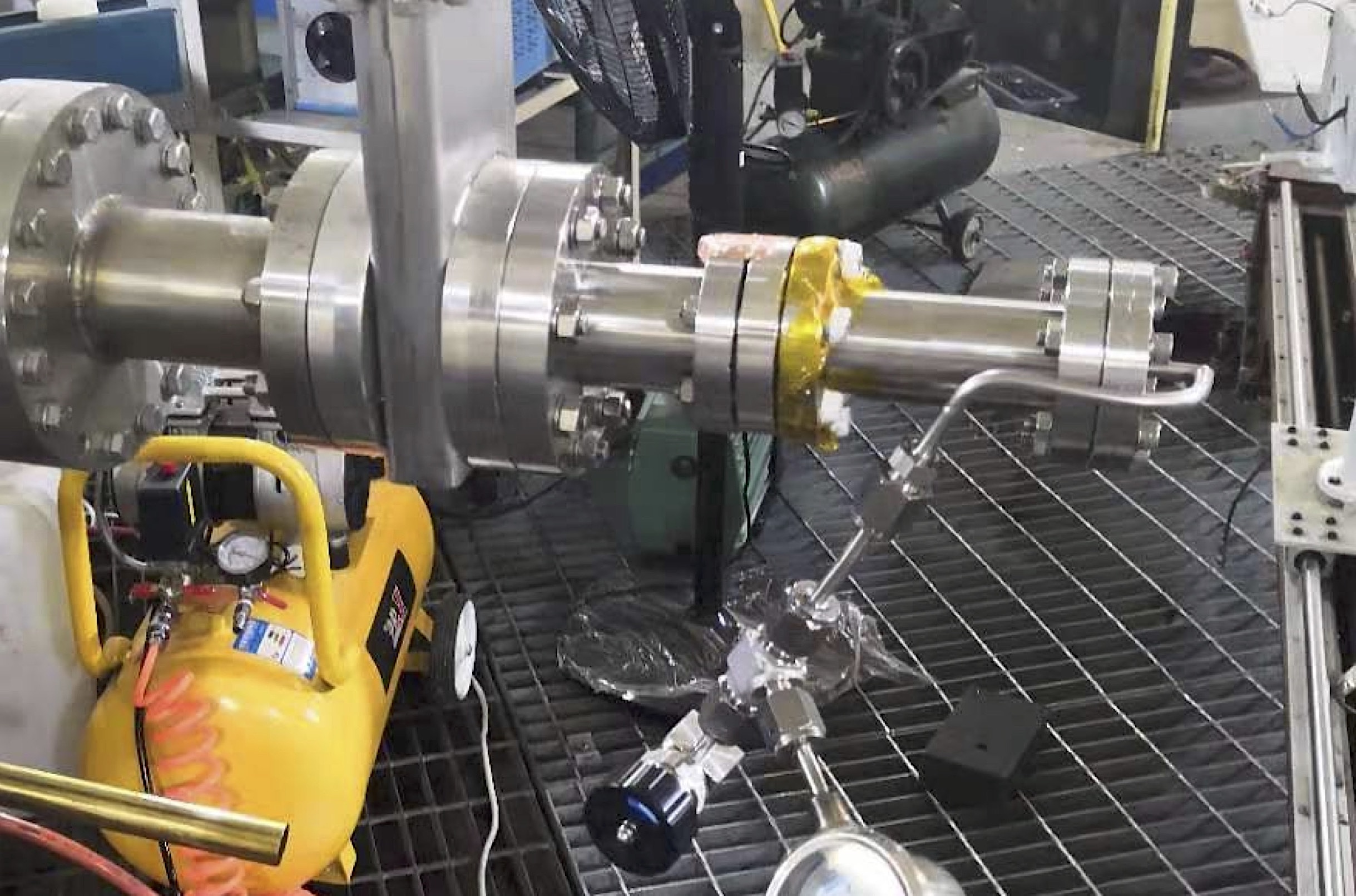}
\caption{Design of $^{\rm{nat}}$Kr target cell.}
\label{fig:chamberdesign}
\end{figure}

(I) We use 20~$\mu$m aluminum~(Al) foil as the bombarding window to separate the target gas and vacuum based on the work in \cite{laura}. The window diameter is set as 10~mm to prevent the foil from breaking due to potentially too large a force on the edge. The
Al foil was tested to withstand a 1.5 bar differential pressure across it.

(II) To measure the proton beam currents, the $^{\rm{nat}}$Kr chamber must be insulated from the upstream parts. Therefore, we use polytetrafluoroethylene~(PTFE) gaskets instead of copper gaskets to seal the CF35 flanges. In addition, the bolts and the flanges are wrapped by Kapton tape, and paper gaskets are put between the stainless steel bolts and the flanges.

(III) The energy deposited by protons on the foil can result in a locally high temperature environment. According to the stopping power of protons in Al provided by PSTAR, 3.4~MeV proton beam could deposit 0.4~MeV in 20~$\mu$m Al~\cite{pstar}. Assuming a proton beam with 2~mm radius and 10~$\mu$A current, the power deposit in the Al foil would have 4~W and the temperature at the center of the Al foil would reach approximately 360~K if the heat dissipates only through conduction, which is still far below the melting temperatures of PTFE or Al.

(IV) To determine the length of the target chamber, we calculate the effective reaction length of the proton beam in $^{\rm{nat}}$Kr using the stopping power of protons in krypton~\cite{pstar}. The estimated reaction length is approximately 5.2~cm. Hence the length of the target chamber is chosen to be 10$\sim$15~cm.

(V) The beam may hit the stainless steel wall of the $^{\rm{nat}}$Kr chamber, producing unexpected radioactive isotopes.
Even though the Coulomb barrier of iron is larger than the proton energy, we put an Al dump, a 5~mm thick pipe with a 5~cm end cap, 
in the $^{\rm{nat}}$Kr chamber to be conservative. The Al dump is wrapped by Kapton tape for electrical insulation.

The two chambers were filled with 1 bar $^{\rm{nat}}$Kr. Two bombardment tests were performed with stable operation on the first chamber for 39 minutes and on the second one for 175 minutes. The average currents of the proton beam were 1.5~$\mu$A and 1.6~$\mu$A for the first and second bombardment respectively.

\section{Production rate measurement with 3.4~MeV protons} \label{sec:measurement}
\begin{figure}[!htb]
\centering
\includegraphics[width = 2.6in]{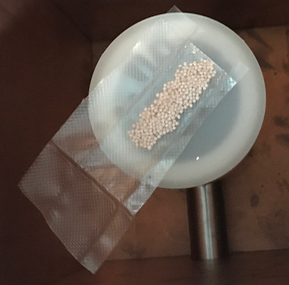}
\caption{Storage of the baked zeolite beads after absorbing rubidium in the de-ionized water solution.}
\label{fig:storage}
\end{figure}

After bombardment, we loosened the bolts of the target cell before further processing to ensure that the rubidium was fully oxidized and in the form of a chemical compound. To study the distribution of $^{83}$Rb production in the bombardment, the first target cell was divided into three parts: the Al window (including the foil and the flange), the Al dump and the CF35-straight tube. Each part was washed with 60-150~ml deionized water separately and 2~g zeolite beads (Merck 2~mm diameter, 0.5~nm molecular sieve) were put into the solution to absorb rubidium in the solution. The solution with beads were heated in a water bath gently at 70$\sim$80$^{\circ}$C until being dried. Then the beads were baked at 300$^{\circ}$C under pumping for further degassing before being stored in sealed plastic bags (Fig.~\ref{fig:storage}). 


We obtained the efficiency of the rubidium transfer by comparing the radioactivity of each part before and after washing, which are 66$\pm$2\% for the Al window, 83$\pm$6\% for the Al dump, 92$\pm$1\% for the CF35-straight tube (Tab.~\ref{tab:transfer}).
\begin{table}[htbp]
  \centering
  \caption{The transfer efficiency for different parts and peaks.}
  \label{tab:transfer}
  \begin{tabular}{l c c c}
    \hline\hline
    Item & {Al Window}		&		{Al dump}		&		CF35-straight tube 		\\
    \hline
    $^{83}$Rb (520)	&	65.7	$\pm$	0.4\%	&	84.2	$\pm$	0.8\%	&	92.3	$\pm$	0.1\%	\\
    $^{83}$Rb (530)	&	65.7	$\pm$	0.7\%		&	84.7	$\pm$	1.0\%	&	92.0	$\pm$	0.1\%	\\
    $^{83}$Rb (553)	&	64.7	$\pm$	1.1\%	&	82.9	$\pm$	1.9\%	&	91.1	$\pm$	0.3\%	\\
    $^{84}$Rb (882)	&	66.9	$\pm$	1.8\%	&	78.7	$\pm$	4.4\%	&	92.2	$\pm$	0.6\%	\\
    $^{86}$Rb (1077)	&	67.8	$\pm$	1.2\%	&	83.3	$\pm$	4.2\%	&	92.5	$\pm$	0.4\%	\\
    Average& 66 $\pm$ 2\% & 83 $\pm$ 6\% &92 $\pm$ 1\%\\
    \hline \hline
  \end{tabular}
\end{table}

\begin{figure}[htbp]
  \centering
  \subfigure[Sample 1 spectrum]{
    \label{fig:samplea}
	\includegraphics[width=0.3\textwidth]{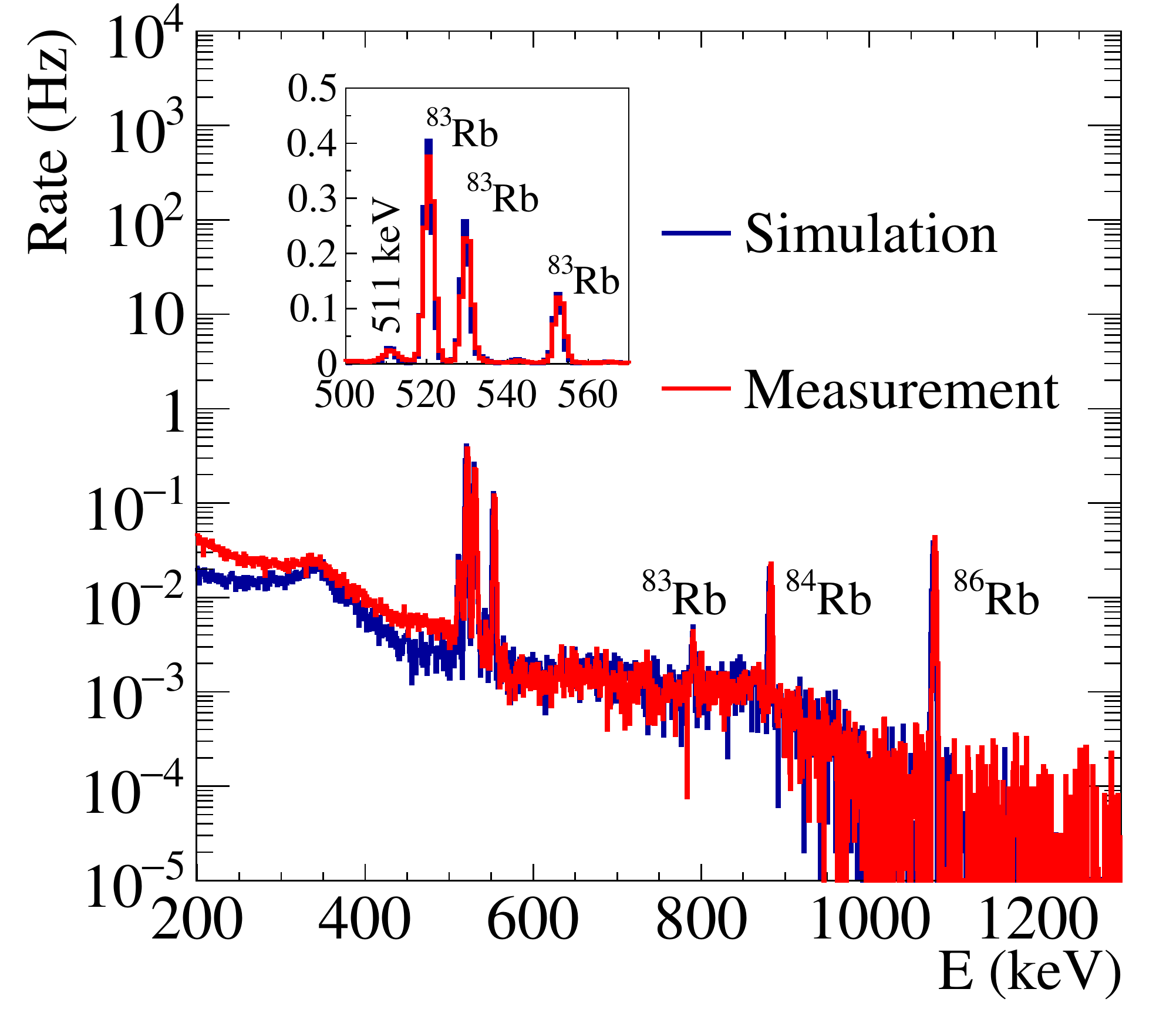}}
  \subfigure[Sample 2 spectrum]{
	\includegraphics[width=0.3\textwidth]{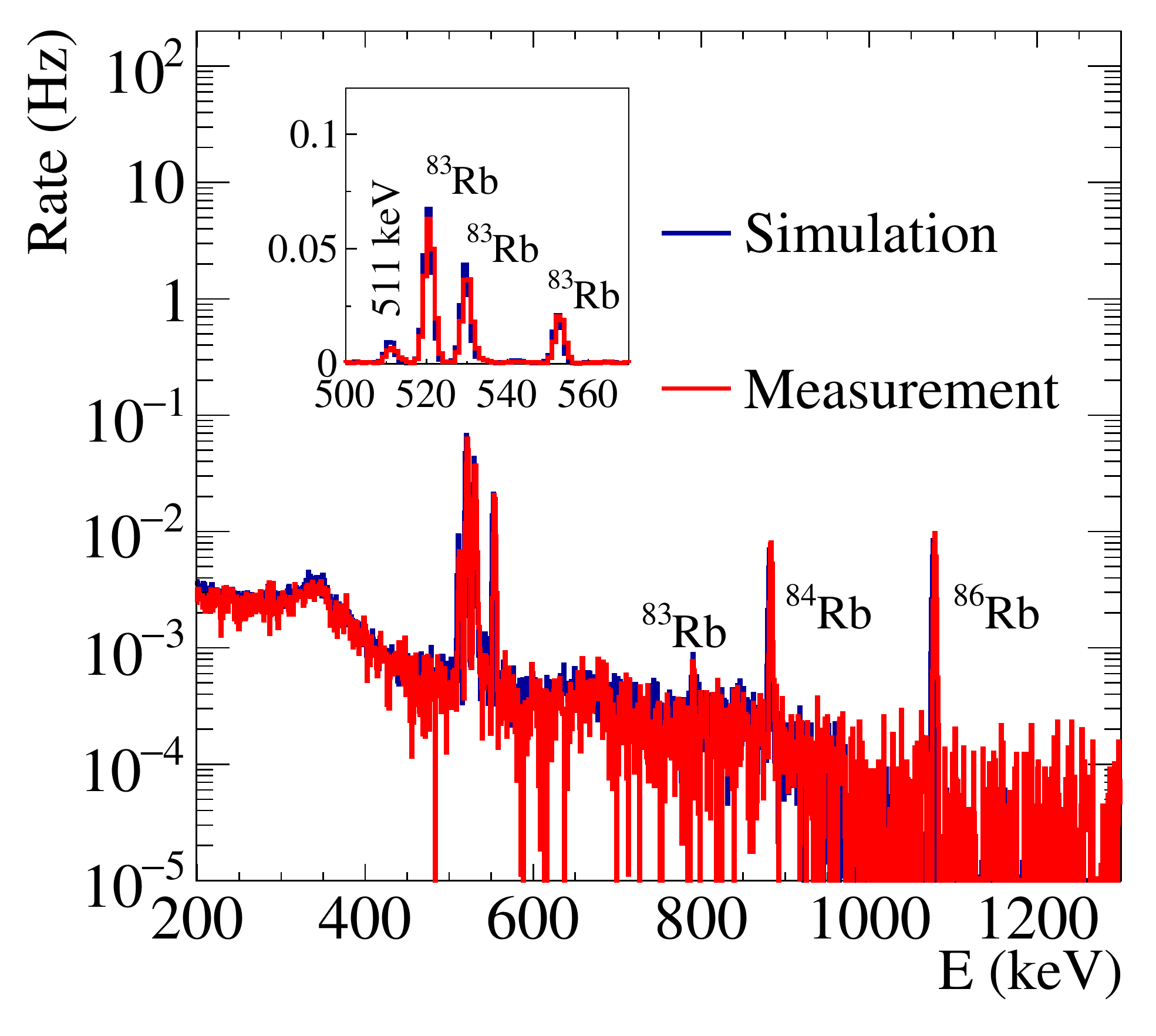}}
  \subfigure[Sample 3 spectrum]{
	\includegraphics[width=0.3\textwidth]{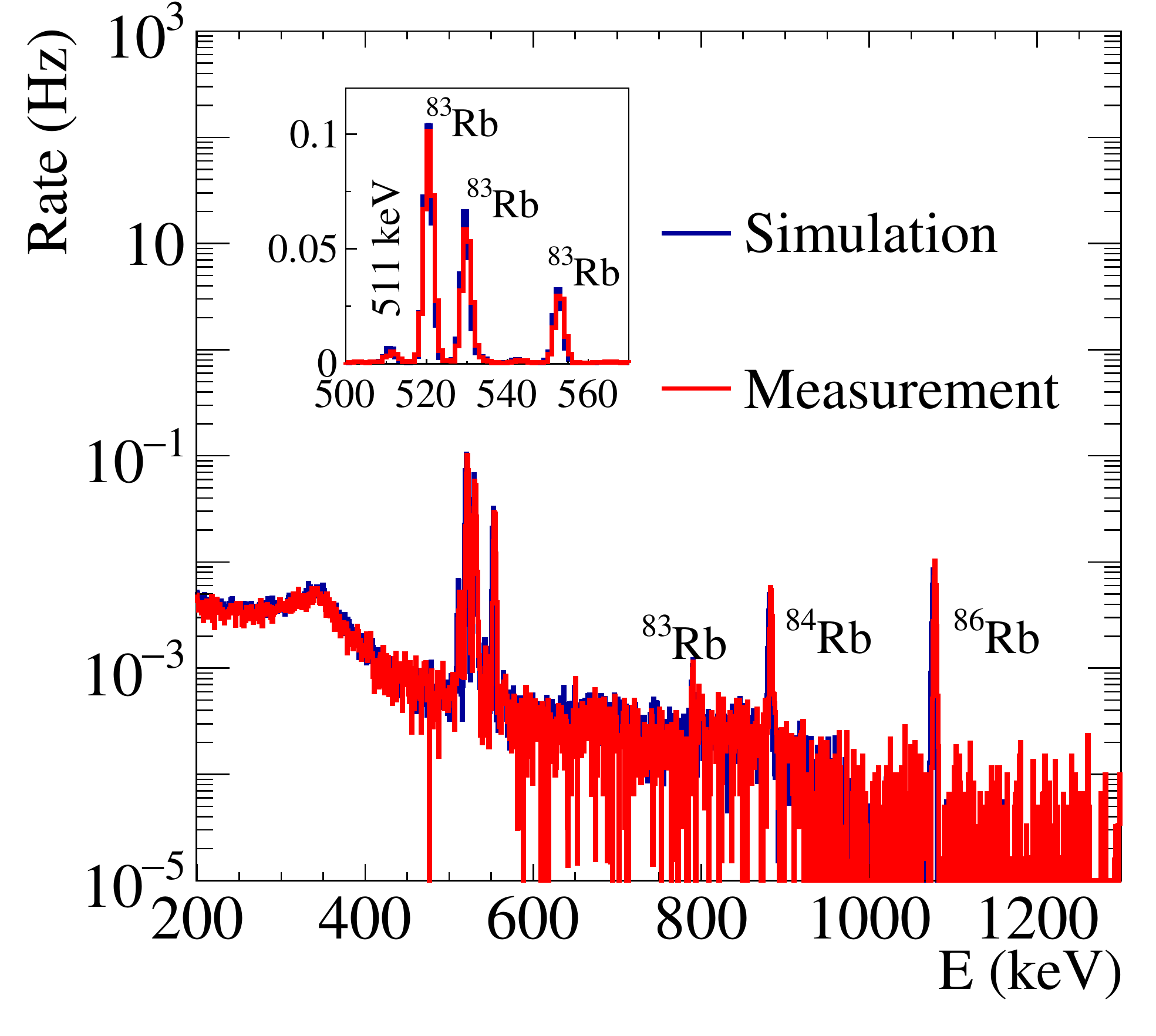}}
  \caption{The comparison between the simulation and the measurement for different zeolite bead samples. Sample 1, sample 2, sample 3 absorbs rubidium from the Al window, the Al dump pipe and the CF35-straight tube respectively. The mismatch of the Compton continuum in sample 1 measurement is likely due to the inaccurate modeling of the sample geometry in the simulation (see Sec.~\ref{sec:production20mev} for details).}
  \label{fig:sample} 
\end{figure}

After being degassed, the zeolite samples were measured by a germanium (Ge) detector at Shanghai Jiao Tong University.  The spectra of the samples are shown in Fig.~\ref{fig:sample} and compared with the GEANT4 Monte Carlo simulation. The radioactivity of each sample is summarized in Tab.~\ref{tab:absolute3}. The systematic uncertainty is dominated by the simulation of geometric detection in the Ge detector, which is 60\% for each simulation. Detailed discussion is presented in Sec.~\ref{sec:production20mev}. The geometric effect may be mitigated if the objects measured are moved farther away above the Ge detector, or if more effort was taken to model the geometric distribution of activity in the zeolite beads. However, our sources produced with 3.4~MeV protons are so weak that the detection efficiency is a more important consideration, and in any case, we did not need to measure the activity to better than the assumed 60\% uncertainty. Compared to the geometric effect, the statistical uncertainty (generally 5\%) and the uncertainty due to rubidium not plating on the surface of the cell are negligible. 

\begin{table}[htbp]
  \centering
    \caption{The radioactivity of the three zeolite samples measured from the Ge detector (only statistical uncertainties applied in this table). }
  \label{tab:absolute3}
  \begin{tabular}{l c c c}
    \hline\hline
    Isotope	&	{Al Window  (Bq)}	&	{Al dump (Bq)}		&	CF35-straight tube (Bq)			\\ \hline
    Time after bombarding	&	40 days  		&	28 days 		&	41 days			\\\hline
    $^{83}$Rb	& 53.4 $\pm$ 2.7 &	9.5	$\pm$ 0.4	&		14.7 $\pm$ 0.6\\
    $^{84}$Rb	& 3.68 $\pm$ 0.14 &	1.35 $\pm$ 0.06	&	0.95 $\pm$ 0.05	\\
    $^{86}$Rb	& 62.6 $\pm$ 1.7 &	15.2 $\pm$ 0.6	&	15.3 $\pm$ 0.6	\\
    \hline \hline
  \end{tabular}
\end{table} 

From the measured zeolite radioactivity and rubidium transfer efficiency, the relative rubidium distribution in the cell (Window : Al dump : CF35-straight tube ) is estimated to be $ 100: (13\pm11):(20\pm17)$ for $^{83}$Rb, $100:(23\pm20):(19\pm16)$ for $^{84}$Rb and $100:(12\pm10):(18\pm15)$ for $^{86}$Rb, respectively (the decays of the isotopes have been considered). The distribution indicates that the rubidium is mainly produced near the Al window, which is as expected because the proton energy is slightly above the threshold of the nuclear reactions and a proton should lose all its energy in approximately 5~cm.

{\color{black}
To determine the total radioactivity generated in the bombardment, the measured radioactivity of each zeolite sample in Tab~\ref{tab:absolute3} is summed and corrected for the half-lives and the transfer efficiencies. The total radioactivity generated in one target cell is $149\pm 89$ Bq for  $^{83}$Rb, $18 \pm 11$ Bq for $^{84}$Rb and $546 \pm 328$ Bq for $^{86}$Rb, respectively. Given the dominance of the Al window in the Rb distribution and relatively simple geometry, we can estimate the total radioactivity directly from the Al window before washing. In the simulation, the Al window is put at the center of the Ge detector, and we assume that the rubidium is uniformly distributed on one side of the Al foil facing to the detector. The results are then further scaled according to the distribution of rubidium among different parts. Table~\ref{tab:double} summarizes the two independent measurements, which are consistent with each other. The average of these two measurements with a reduced uncertainty is taken to calculate the thick target yield. Considering the charge of proton used is $3.5\times 10^{-3}$~C, we obtain the thick target yields for the production of rubidium isotopes by the 3~MeV (effective energy after losing energy in the Al foil) protons bombarding $\rm^{nat}$Kr, as shown in Tab.~\ref{tab:double}.

\begin{table}[htbp]
  \centering
    \caption{The total generated radioactivities and thick target yields for the production of rubidium isotopes with the 3~MeV protons bombarding $\rm^{nat}$Kr determined by the spectra of zeolite samples and Al window before washing. See context for more details.}
  \label{tab:double}
  \begin{tabular}{l c c c c}
    \hline\hline
    Isotope	&	Zeolite (Bq) & Al Window (Bq) & Average (Bq)	&	Thick target yield (MBq/C)			\\ \hline
    $^{83}$Rb	&	149	$\pm$ 89	&	136	$\pm$ 81	&		142 $\pm$ 60  & 0.041 $\pm$ 0.017 \\
    $^{84}$Rb	&	18	$\pm$ 11	&	15	$\pm$ 9	&	16 $\pm$ 7 & 0.005 $\pm$ 0.002	\\
    $^{86}$Rb	&	546	$\pm$ 328	&	556	$\pm$ 336		& 551 $\pm$ 234 & 0.16 $\pm$ 0.07\\
    \hline \hline
  \end{tabular}
\end{table}

\begin{figure}[htbp]
  \centering
  \subfigure[]{
    \label{fig:rb83th}
    \includegraphics[width=2.6in]{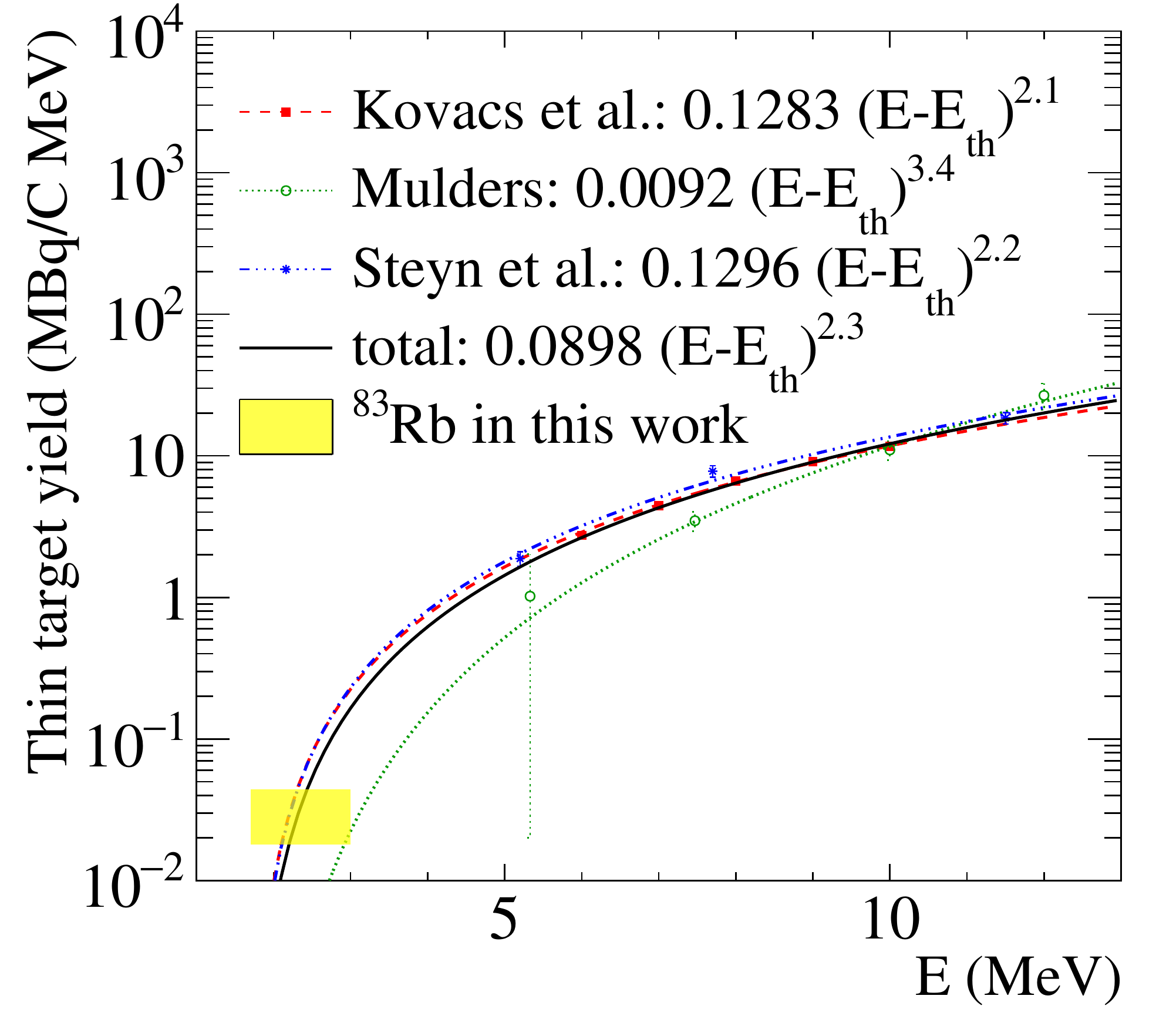}
  }
  \subfigure[]{
    \label{fig:rb86th}
    \includegraphics[width=2.6in]{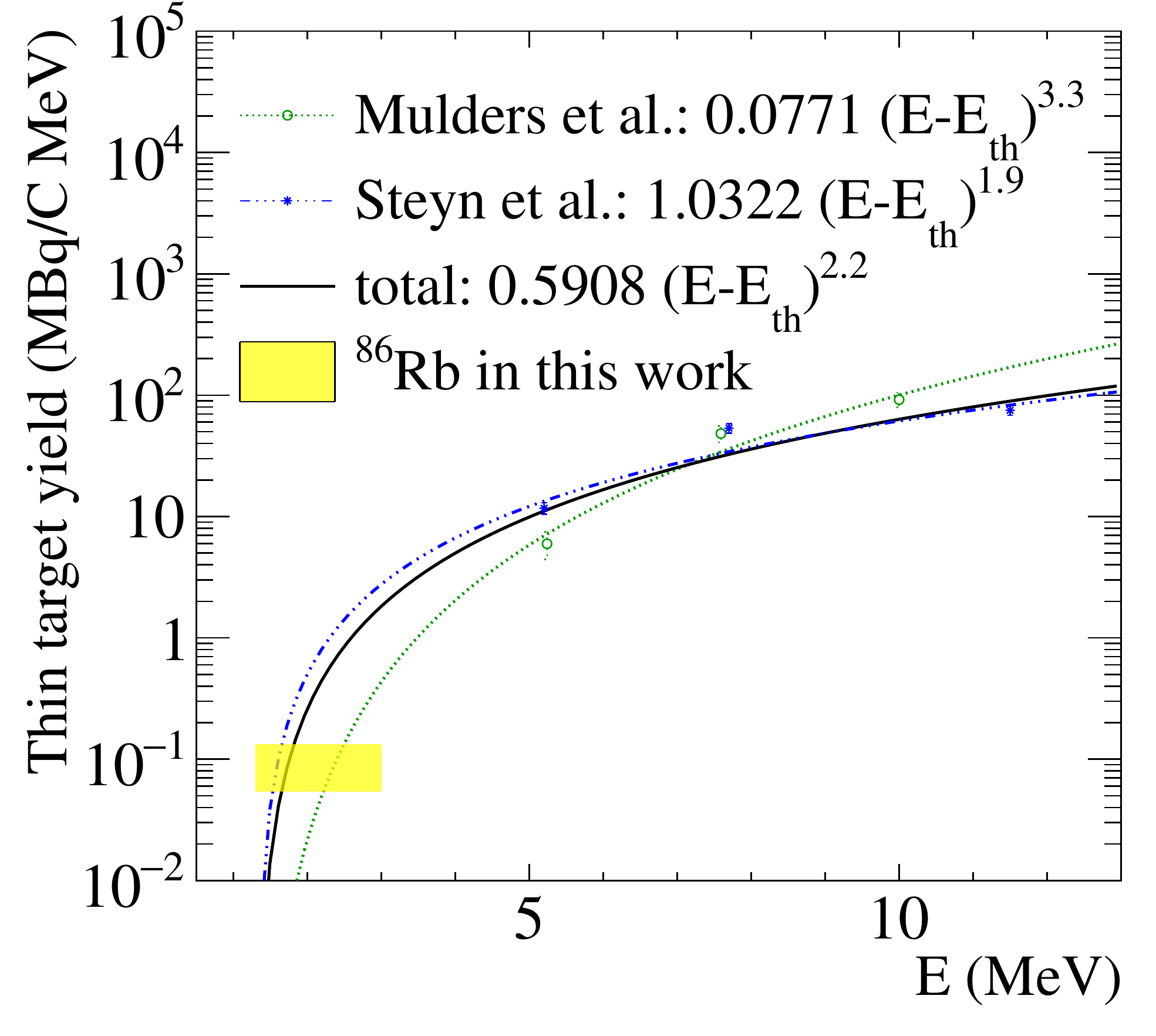}
  }
  \caption{The thin target yield comparison between this work and extensions from previous works (``total'' means take all the previous data into account) for $^{83}$Rb (left) and $^{86}$Rb (right)~\cite{natKrRb,natKrRb1,natKrRb2}. The energy threshold, $E_{\rm th}$, is 1.7~MeV for the $^{83}$Kr(p,n)$^{83}$Rb reaction and 1.3~MeV for the $^{86}$Kr(p,n)$^{86}$Rb.
   }
  \label{fig:thcomparison} 
  \end{figure}

Our result of $^{83}$Rb and $^{86}$Rb production is converted to thin target yields by dividing the thick target yields by the energy window above the known reaction energy thresholds, and compared with a phenomenological extrapolation from the measured high energy yields~\cite{natKrRb, natKrRb1, natKrRb2}, as shown in Fig.~\ref{fig:thcomparison}. According to L. D. Landau's theory, the cross section near the reaction threshold is proportional to the velocity of the proton~\cite{qmlandau}.  However, the high energy yields do not fit this simple theoretical model. The extrapolation function is chosen as some power of the velocity with the power being a free parameter to be fitted. 

In the production, we observed a small amount of $^{84}$Rb unexpectedly. The protons are accelerated with a tandem pelletron (1.7~MV, Model 5SDH, National Electrostatics Corp.)~\cite{tandem}, the highest energy of incoming protons is limited to 3.4~MeV with an uncertainty of 1~keV, which is below the theoretical threshold 3.46~MeV~\cite{natKrRb,nndc84} of the $^{84}$Kr(p,n)$^{84}$Rb reaction. The $^{84}$Kr(p,n)$^{84}$Rb reaction is unlikely to happen in this bombardment theoretically. 
}

\section{Production with 20 MeV protons} \label{sec:production20mev}
Recently, a new proton facility- Chinese ADS Front End demo linac (CAFE) was built at the Institute of Modern Physics, Chinese Academy of Sciences, with energy up to 25~MeV~\cite{cafe,ciads}. As one of the first users, we conducted another test with the 20~MeV proton beam bombarding 1.1~bar $\rm^{nat}$Kr. The total exposure to the protons was 9.7~$\mu$Ah (0.035~C). 

We preserved the previous design of the target cell and added extra water cooling for the 20~MeV proton bombardment.
In this test, the main concern is the heat loads on the target cell instead of the Al window, because each proton deposits only 0.11~MeV in the Al foil according to the stopping power on PSTAR~\cite{pstar}. 
To stop the 20~MeV protons with 1~$\mu$A average currents, the heat gain on the back of the target cell is 20~W. The Al dump is cooled by room temperature water with a flux up to 400~cm$^3$/s.

Multiple $^{83}$Rb/$^{83m}$Kr sources from several kilo to mega Becquerel were obtained in the processing procedure. The Al window and the target chamber were washed by deionized water separately (the chamber was washed three times). 
The radioactivity ratio of the final zeolite samples is $\rm 1^{st}:2^{nd}:3^{rd}:window=1:0.050:0.0032:0.052$. The transferring efficiency of the main target chamber in one wash was determined to be 90\% as before. The strongest $^{83}$Rb/$^{83m}$Kr source obtained is approximately 10~MBq.

\begin{figure}[!htb]
  \centering
  \subfigure[]{
    \includegraphics[width=0.3\textwidth]{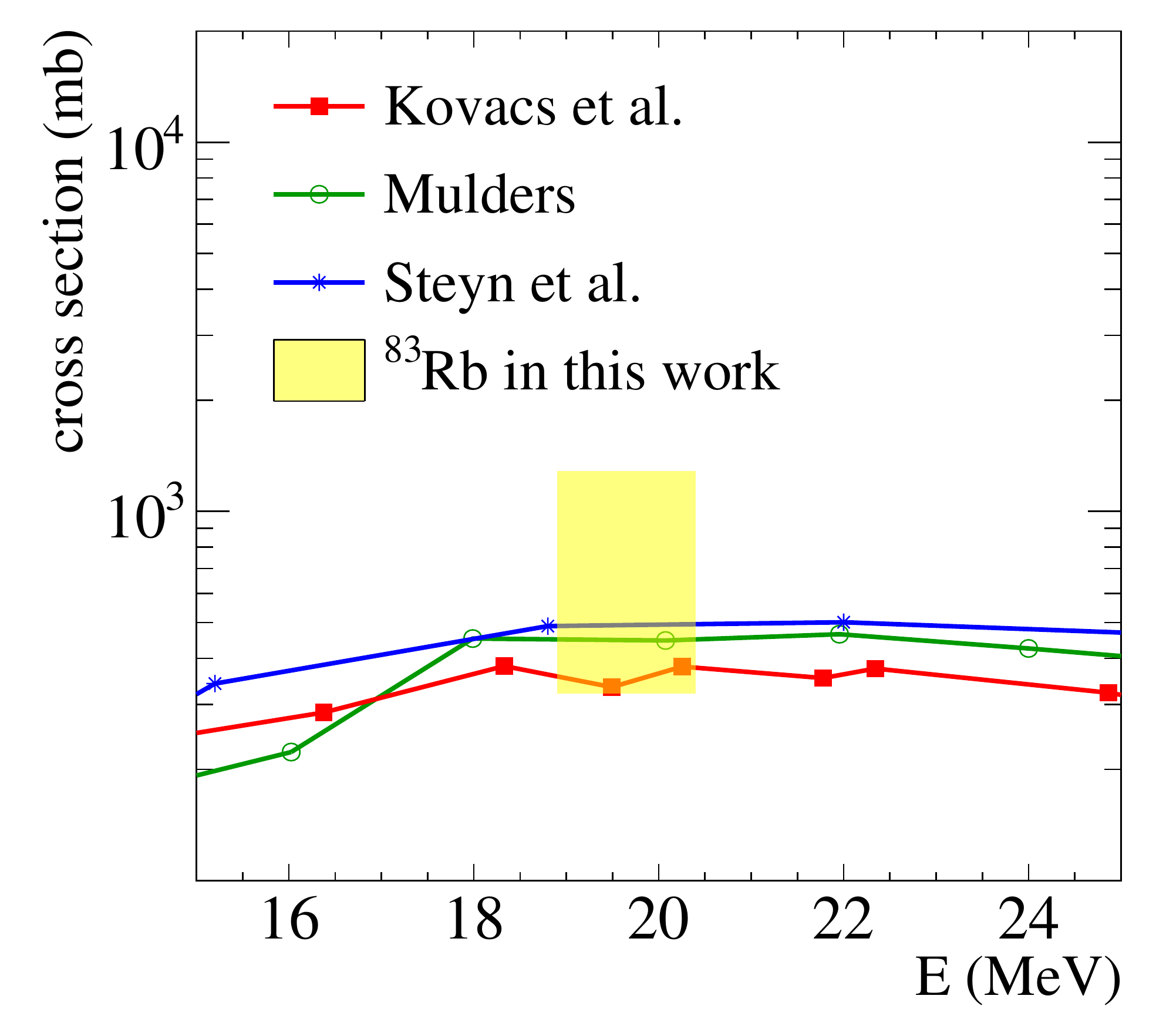}}
  \subfigure[]{
    \includegraphics[width=0.3\textwidth]{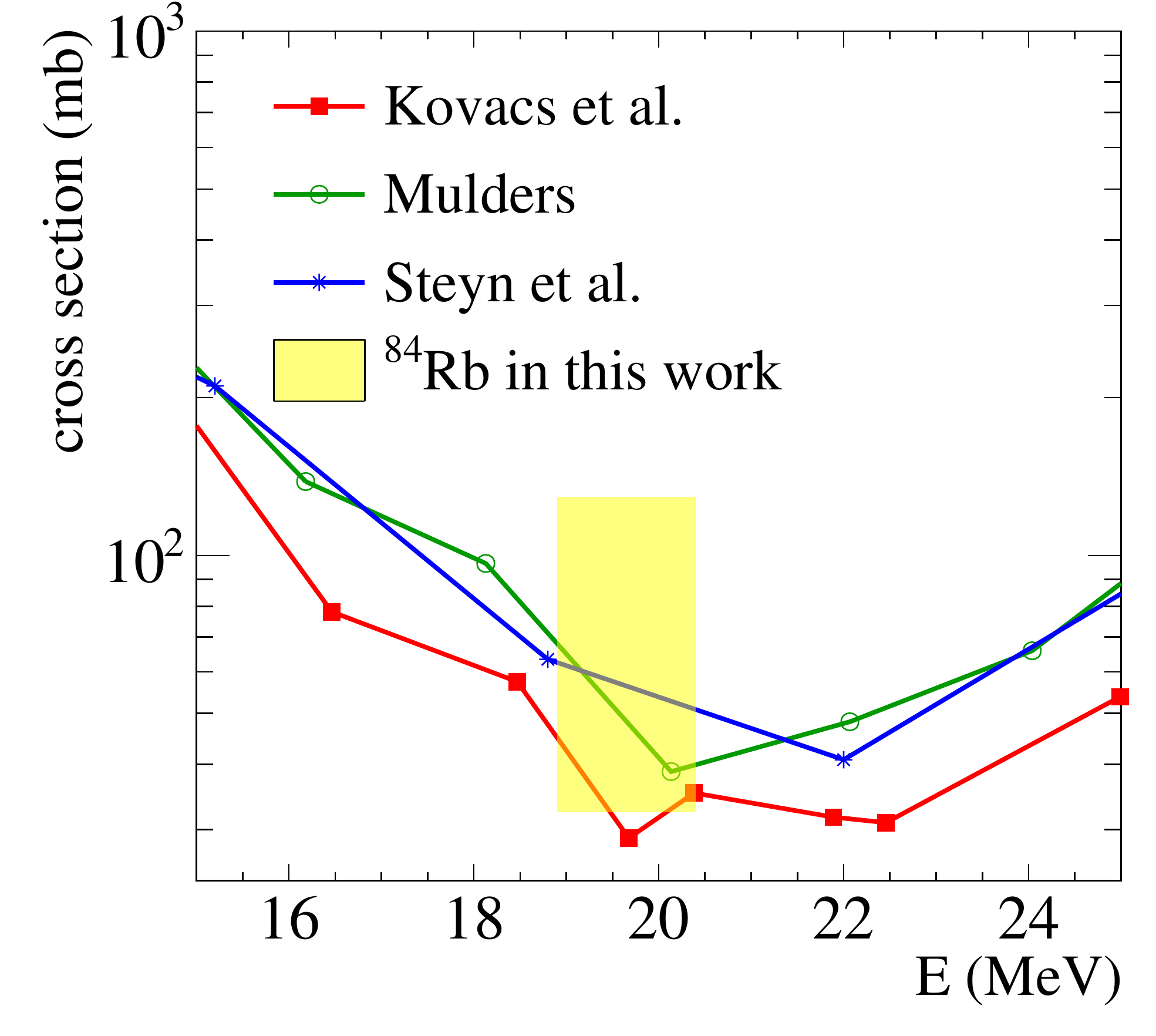}}
  \subfigure[]{
	\includegraphics[width=0.3\textwidth]{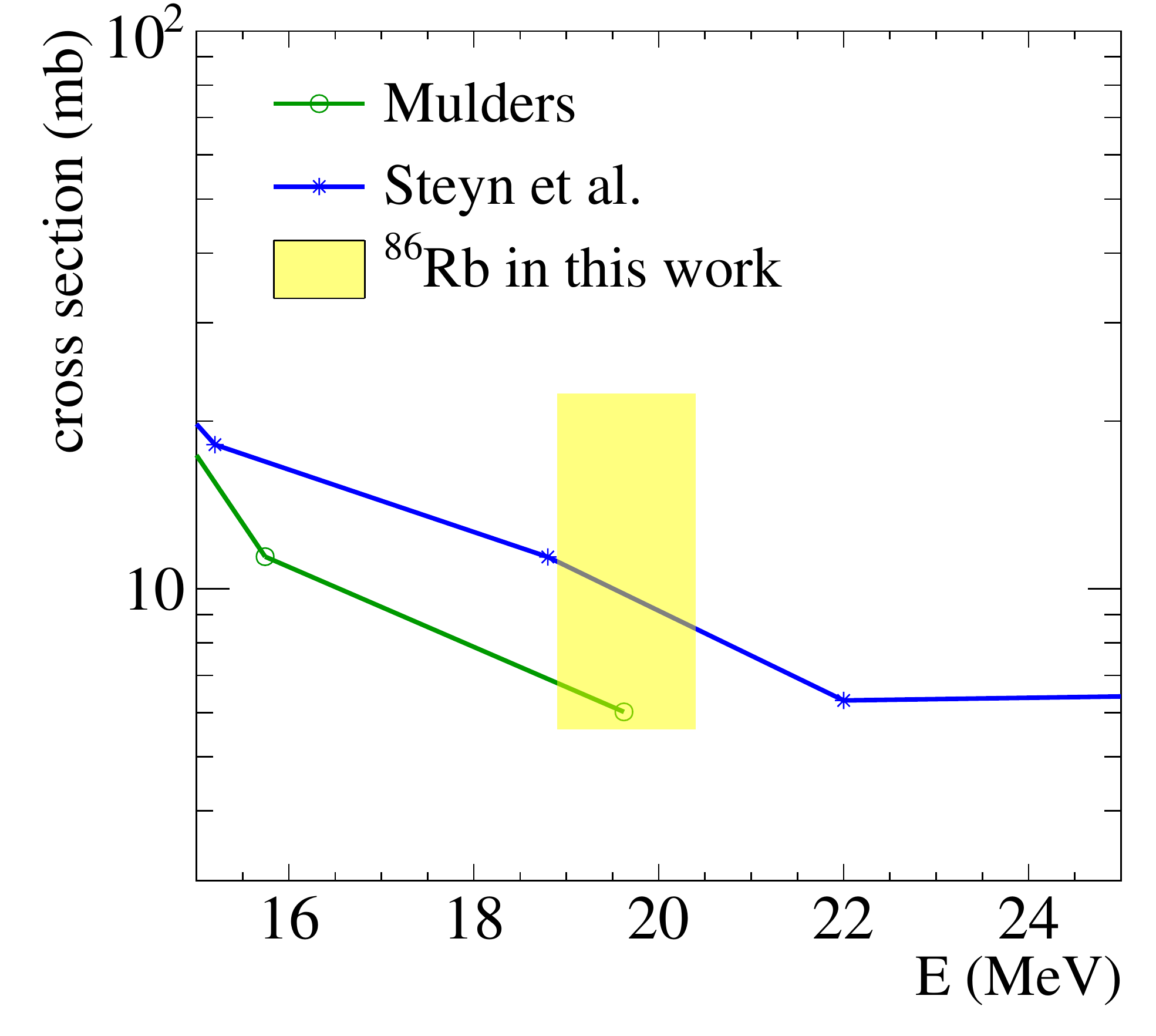}}
  \caption{Comparison of the (a) $^{\rm{nat}}$Kr(p, xn)$^{83}$Rb, (b) $^{\rm{nat}}$Kr(p, xn)$^{84}$Rb and (c) $^{\rm{nat}}$Kr(p, xn)$^{86}$Rb cross sections at 20~MeV  between this work and previous measurements~\cite{natKrRb,natKrRb1,natKrRb2}.} 
  \label{fig:xsec20mev} 
\end{figure}

The cross-section ($\sigma$) of $^{\rm{nat}}$Kr(p, xn)$^{83}$Rb, $^{\rm{nat}}$Kr(p, xn)$^{84}$Rb and $^{\rm{nat}}$Kr(p, xn)$^{86}$Rb at 20~MeV measured by the Ge detector is consistent with the previous measurements, as shown in Fig.~\ref{fig:xsec20mev}. The initial proton energy calculated with the time of incident protons flying through a 2.47~m vacuum chamber is $20.37\pm0.03$~MeV. According to the stopping power on PSTAR, in the 25 cm target cell filled with 1.1 bar krypton gas, the proton energy loss is 1.4 MeV~\cite{pstar}. This energy loss dominates the energy spread in Fig.~\ref{fig:xsec20mev}. We validate the systematic uncertainties by measuring the detecting efficiency of a millimeter-scale source. Compared to the rubidium sources, the cylindrical shaped calibration source with a 3~mm radius and 6~mm height is small enough to be regarded as a point source. The typical size difference among the $^{83}$Rb/$^{83m}$Kr sources is 3~cm. According to the measurements with the calibration source, a 3~cm horizontal deviation to the surface center reduces the detecting efficiency to 60\% and a 3~cm vertical deviation to 40\%. Therefore, the systematic uncertainties are set to 60\% in Fig.~\ref{fig:xsec20mev}.

\section{Injection of $^{83m}$Kr in PandaX-II detector} \label{sec:injection}
We used 1~g zeolite carrying 30~Bq (60\% uncertainty) $^{83}$Rb for the detector injection test. In order to stop zeolite from escaping the source chamber, we put a 0.2~$\mu$m membrane filter (Merck, FGLP01300), which has an upper limit of rubidium leakage as $1.3\times10^{-10}$~\% per hour (2.4~$\mu$Bq/h for a 1.8~MBq $^{83}$Rb/$^{83m}$Kr source) \cite{vhannen}. 
The chamber with zeolite filled was first pumped separately for 60~h at 80$^{\circ}$C to reach a vacuum of $5.8\times10^{-4}$~Pa, then 20~L gas xenon was injected into the chamber to mix with $^{83m}$Kr. The mixed gas was circulated and purified by a getter (PS4-MT50-R-2) for 24~h. The source chamber was then connected to the PandaX-II detector for 12~h. 

The decay from $^{83m}$Kr to the ground state is mainly through internal conversion electrons ($\tau=1.83$~h). The direct decay mode from $41.5$~keV to the ground state is suppressed due to the spin difference. Instead, the decay occurs in two transitions of $32.1$~keV and $9.4$~keV respectively, with an intervening half-life of 154 ns, as shown in Tab.~\ref{tab:ce83mkr}~\cite{ce83mkr}. If the detector responses quickly enough, we should be able to see the two transitions ($32.1$~keV and $9.4$~keV). However, the de-excitation of the Xe$_{2}^{*}$ has a lifetime of 35~ns at 40~keV~\cite{deexcitation}, and the electronics limit the width of the pulse to the level of 100~ns. Hence only part of the scintillation signals (S1s) of the two peaks can be separated. For the ionization signals (S2s), the time resolution of our detector is limited to several microseconds, which makes it almost impossible to separate the two transitions. We did not see any event with two separate S2s in the $5\times10^{5}$ $^{83m}$Kr events collected.

\begin{table}[htb!]
  \centering
  \caption{Decay channels of the $^{83m}$Kr isotope \cite{ce83mkr}.}
  \label{tab:ce83mkr}
  \begin{tabular}{l l  l}
    \hline\hline
    Transition energy	&	Decay mode		&		Branching	ratio	\\    \hline
    32.1~keV & e(30~keV)+e(2~keV) &76\%	\\
    &			e(18~keV)+e(10~keV)+2e(2~keV)&9\%\\
    &			e(18~keV)+X(12~keV)+2e(2~keV)&15\%\\
    9.4~keV &		e(7.6~keV)+e(1.8~keV) &95\%	\\
    &		$\gamma$						&5\%\\
    \hline \hline
  \end{tabular}
\end{table}

\begin{figure}[!htb]
  \centering
  \includegraphics[width=4in]{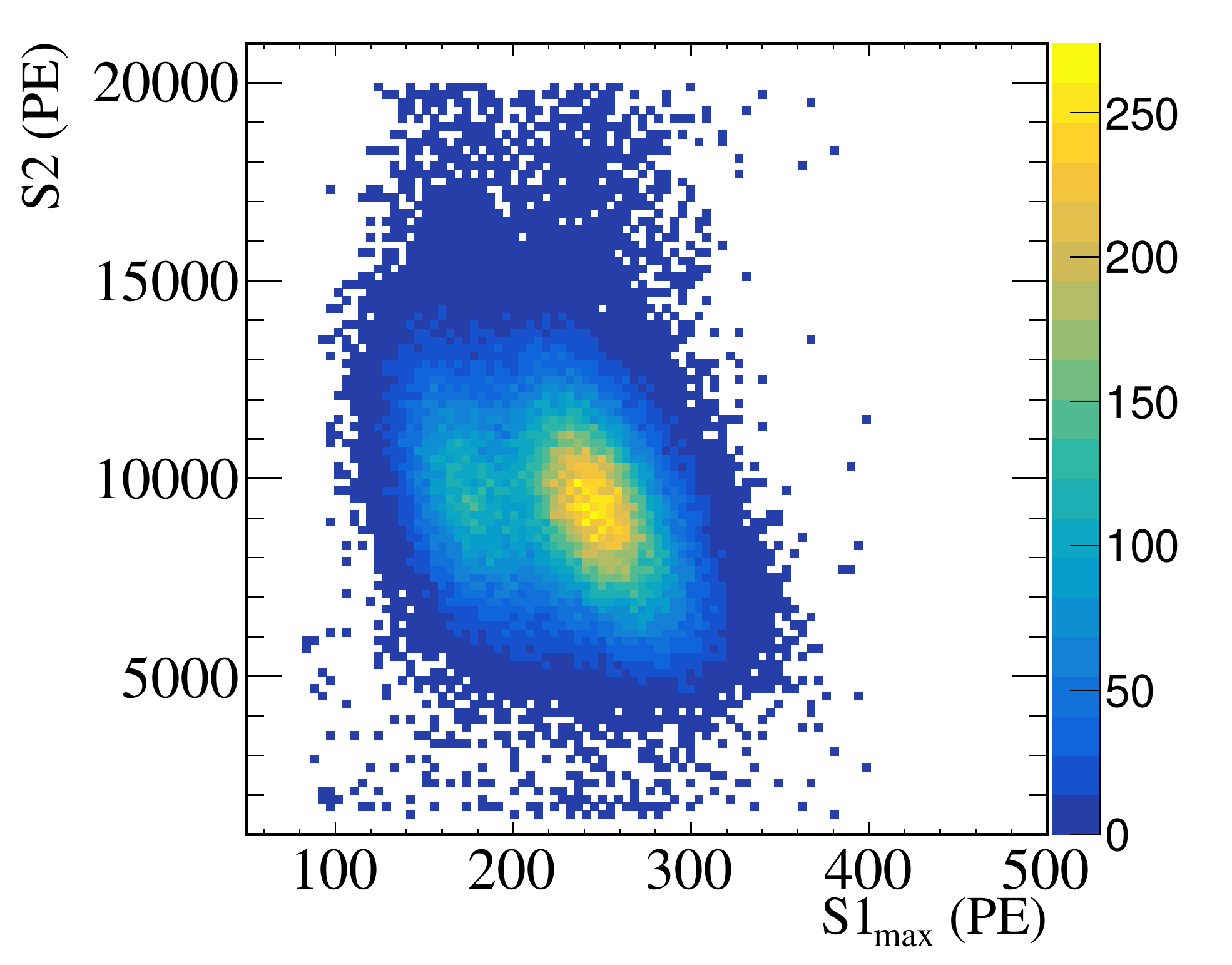}
  \caption{S2 vs S1$\rm_{max}$ by applying the data chain for the WIMP search runs and the color bar represents the counting.}
  \label{fig:original} 
\end{figure}

\begin{figure}[!htb]
  \centering
  \subfigure[]{
    \includegraphics[width=2.5in]{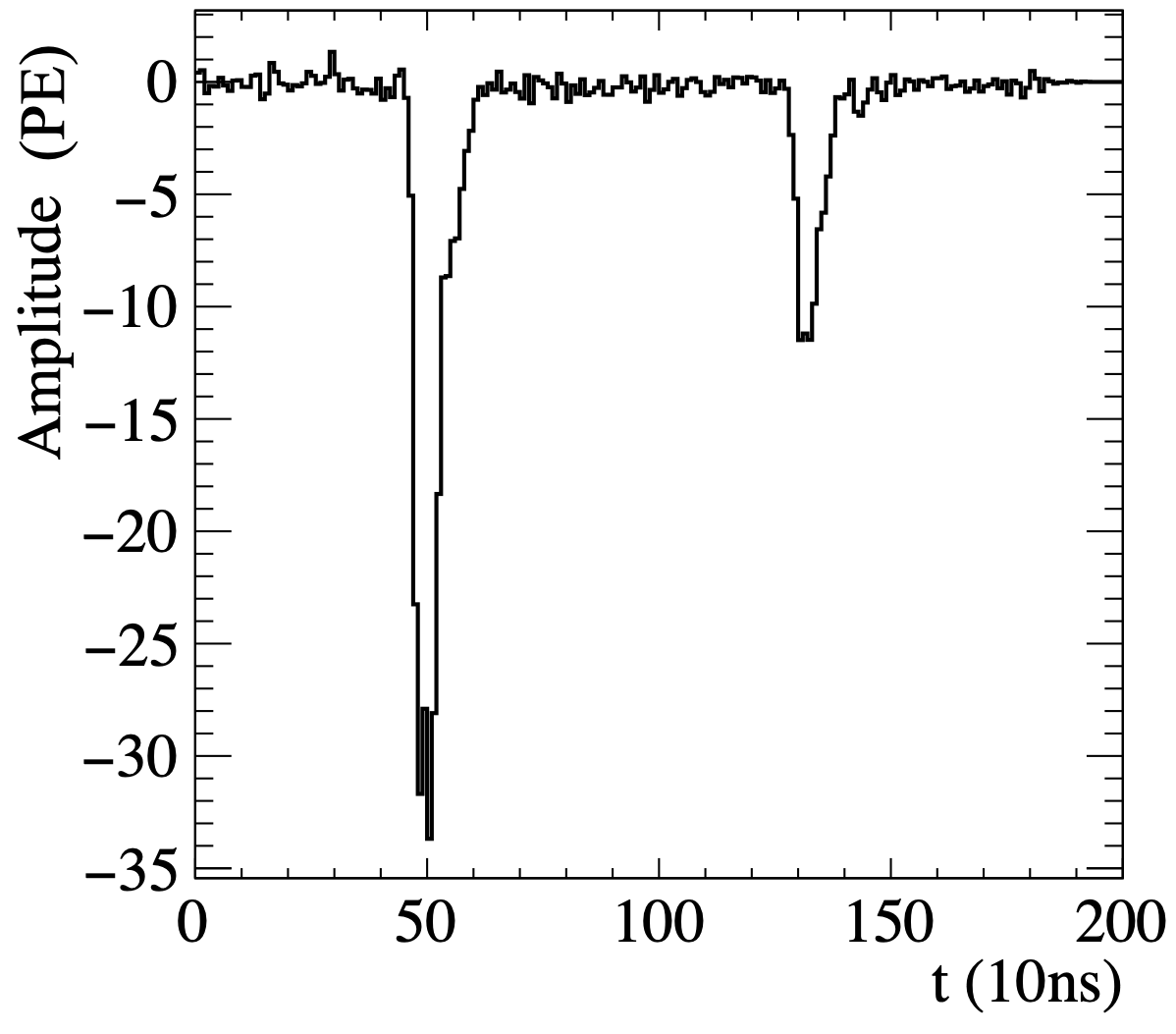}}
  \hspace{0.5in}
  \subfigure[]{
    \includegraphics[width=2.5in]{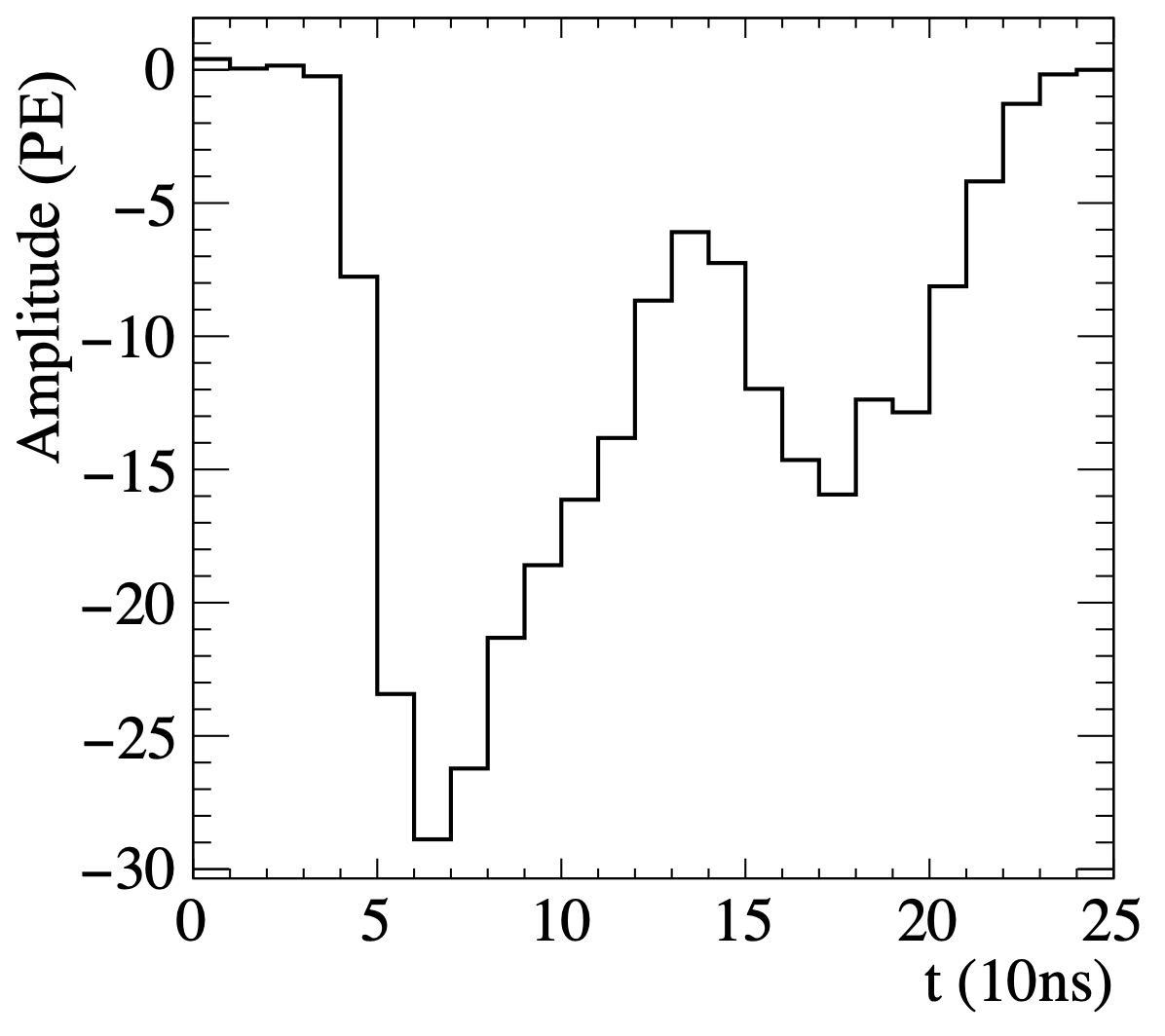}}  
  \caption{(a) The waveform with the two S1s separated.  (b) The waveform with the two S1s mixed.}
  \label{fig:s1}
\end{figure}

\begin{figure}[!htb]
  \centering
  \subfigure[]{
	\label{fig:9.4hl}
    \includegraphics[width=2.6in]{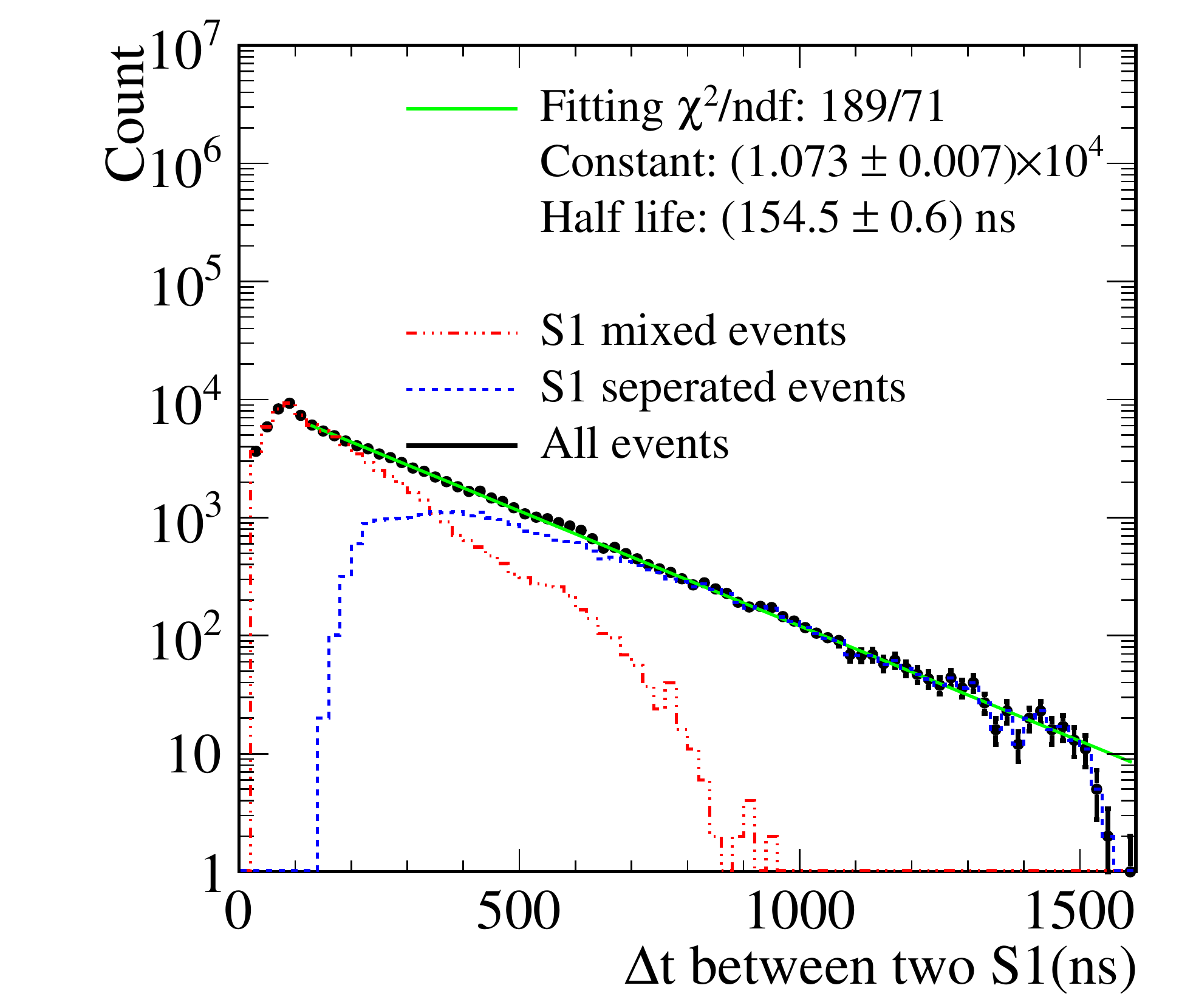}}
  \subfigure[]{
    \label{fig:41.5peak}
	\includegraphics[width=2.6in]{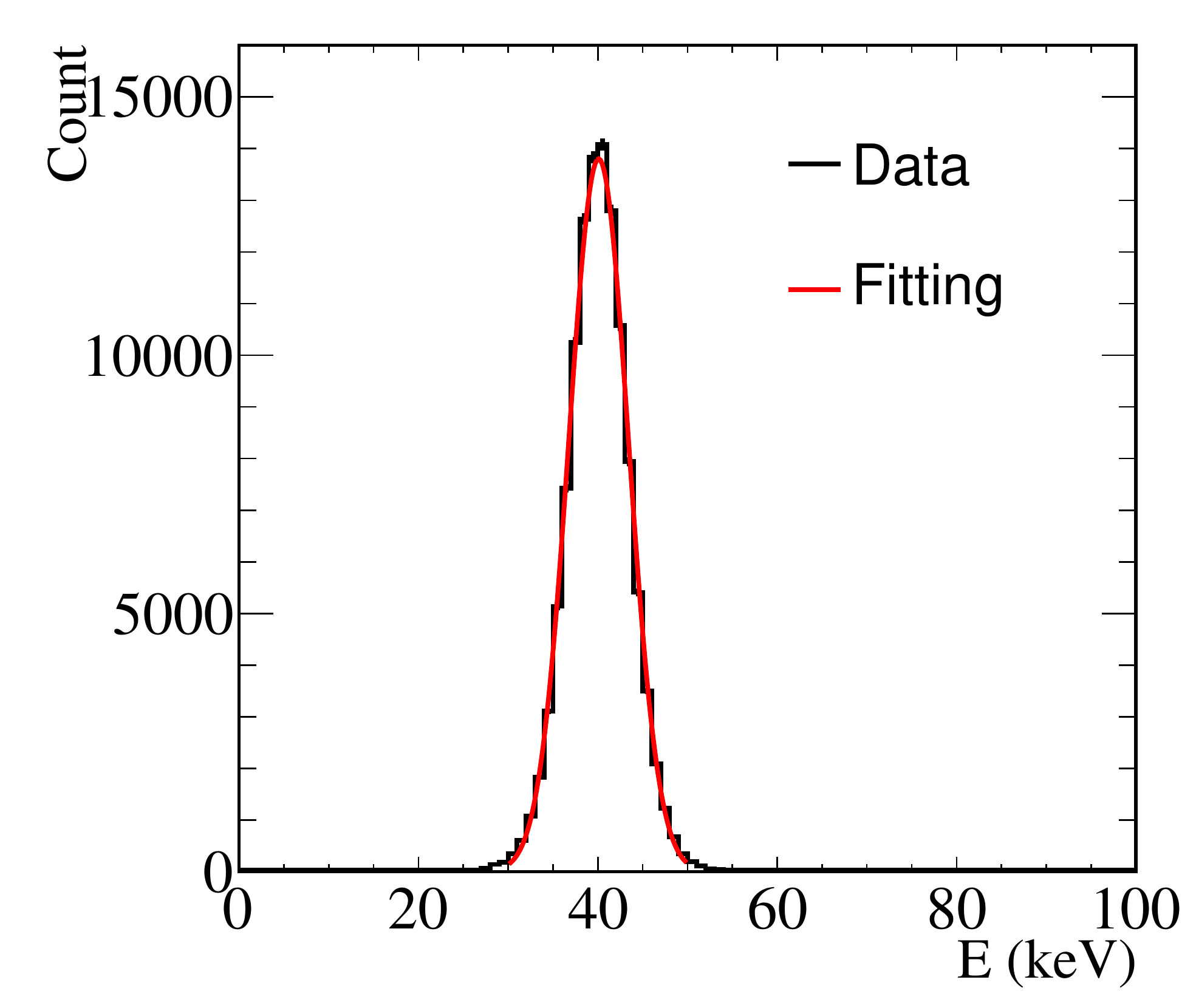}}  
  \caption{(a) The fit of the half-life for the first excited state of $^{83m}$Kr.  (b) The energy spectrum of $^{83m}$Kr with background subtracted.}
  \label{fig:e}
\end{figure}

Figure~\ref{fig:original} shows the response of $^{83m}$Kr in the PandaX-II detector after the data process chain used in \cite{main1,main2}, where we can see two S1 peaks from $^{83m}$Kr with one S2 peak. 
The S1 waveform with the two transitions separated or mixed is shown in Fig.~\ref{fig:s1}. The time interval of the two S1s, if well separated, could be used to fit the half-life of the first excited state of $^{83m}$Kr as shown in Fig.~\ref{fig:9.4hl}. From fitting the tail with $\Delta t$ larger than 120~ns, we obtained a half-life of $154.5\pm0.6$~ns, which is consistent with the theoretical value. 

Figure.~\ref{fig:41.5peak} shows the reconstructed energy spectrum with the sum of two sequential S1s and one S2, which agrees with the $^{83m}$Kr characteristic energy peak. The energy resolution for $^{83m}$Kr is 8.0\%, the mean of the peak $E_0$ is 40.8~keV. The fitted mean value is smaller than that provided by NNDC (41.5~keV) partly due to the baseline suppression threshold of the PandaX-II data-acquiring system. Meanwhile, the detector response model could be further tuned according to this calibration data~\cite{pandax2nr}.

\section{Conclusions}
\label{conclusion}
We report a successful production of $\rm ^{83}Rb/^{83m}Kr$ with 3.4 MeV proton beam provided by the China Institute of Atomic Energy. The production rate is measured to be $0.041\pm0.017$~MBq/C for the $^{83}$Kr(p,n)$^{83}$Rb process, which is the first experimental data ever reported for such low energy proton beams in the world. Another production attempt was performed with the recently available 20 MeV proton beam at the Institute of Modern Physics, Chinese Academy of Sciences, which was one of the first applications on this proton facility. The produced $\rm ^{83m}Kr$ source has been successfully injected into the PandaX-II liquid xenon detector and yielded enough statistics for detector calibration.

\section*{Acknowledgments}
This project is supported in part by grants from National Science Foundation of China (Nos. 11525522, 11775141 and 11755001), the Zhiyuan Scholar Program (No. ZIRC2018-01), the Double First Class Plan of the Shanghai Jiao Tong University and a grant from the Ministry of Science and Technology of China (No. 2016YFA0400301). We thank the Office of Science and Technology, Shanghai Municipal Government (No. 11DZ2260700, No. 16DZ2260200, No. 18JC1410200) for important support. We also thank the sponsorship from the Chinese Academy of Sciences Center for Excellence in Particle Physics (CCEPP), Hongwen Foundation in Hong Kong, and Tencent Foundation in China. Finally, we thank the CJPL administration and the Yalong River Hydropower Development Company Ltd. for indispensable logistical support and other help.

\bibliographystyle{apsrev4-1}
\bibliography{Kr83m_prc.bib}

\end{document}